\documentstyle[pra,aps,multicol,epsfig]{revtex}

\newcommand{\be}[1]{\begin{equation}}
\newcommand{\ee}[1]{\end{equation}}

\begin{document}

\title{Modulational instability in nonlocal nonlinear Kerr media}

\author{Wieslaw Krolikowski}
\address{Australian Photonics Cooperative Research Centre,
Laser Physics Centre,\\
Research School of Physical Sciences and Engineering, \\
Australian National University, Canberra ACT 0200, Australia.}

\author{Ole Bang}
\address{Department of Informatics and Mathematical Modelling, \\
Technical University of Denmark, DK-2800 Kongens Lyngby, Denmark.}

\author{Jens Juul Rasmussen}
\address{Ris\o\ National Laboratory, Optics and Fluid Dynamics Department, \\
OFD - 128, P.O. Box 49, DK-4000 Roskilde, Denmark.}

\author{John Wyller}
\address{Department of Mathematical Sciences, Agricultural University of 
Norway \\ P.O. Box 5065, N-1432 \AA s, Norway.}

\maketitle

\begin{abstract}
We study modulational instability (MI) of plane waves in nonlocal nonlinear 
Kerr media. For a focusing nonlinearity we show that, although the nonlocality 
tends to suppress MI, it can never remove it completely, irrespectively of the 
particular profile of the nonlocal response function. 
For a defocusing nonlinearity the stability properties depend sensitively on
the response function profile: for a smooth profile (e.g., a Gaussian) plane waves are always stable, but MI may occur for a rectangular response.  
We also find that the reduced model for a weak nonlocality predicts MI in defocusing  media for arbitrary response profiles, as long as the intensity exceeds a certain critical value. However, it appears that 
this regime of MI is beyond the validity of the 
reduced model, if it is to represent the weakly nonlocal limit of a general
nonlocal nonlinearity, as in optics and the theory of Bose-Einstein condensates.
\end{abstract}
\begin{multicols}{2}

\section{Introduction}

Modulational instability (MI) constitutes one of the most fundamental effects 
associated with wave propagation in nonlinear media. 
It signifies the exponential growth of a weak perturbation of the amplitude 
of the wave as it propagates.  
The gain leads to amplification of sidebands, which breaks up the otherwise 
uniform wave front and generates fine localized structures (filamentation).  
Thus it may act as a precursor for the formation of bright spatial solitons.
Conversely the generation of dark spatial solitons requires the absence of
MI of the constant intensity background.

The phenomena of MI has been identified and studied in various physical 
systems, such as fluids \cite{mi-fluid}, plasma \cite{mi-plasma}, nonlinear 
optics \cite{mi-nlo1,mi-nlo2}, discrete nonlinear systems, such as molecular chains 
\cite{discrete-mi-molchain} and Fermi-resonant interfaces and waveguide arrays \cite{discrete-mi-array}, etc.  
It has been shown that MI is strongly affected by various mechanisms present 
in nonlinear systems, such as higher order dispersive terms in the case of
optical pulses \cite{mi-dispersive}, saturation of the nonlinearity 
\cite{mi-saturation}, and coherence properties of optical beams 
\cite{mi-coherence}.  

In this work we study the MI of plane waves propagating in a nonlinear 
Kerr type medium with a nonlinearity (the refractive index change 
$\Delta n$, in nonlinear optics) that is a nonlocal function of the 
incident field.  
We consider a phenomenological model, in which the nonlocal nonlinearity, 
induced by a wave (e.g. an optical beam) with the intensity $I(x,z)$, can
be represented in general form as
\begin{equation}
  \label{nonlocal}
  \Delta n(I)= s\int_{-\infty}^{\infty}R(x'-x)I(x',z)dx',
\end{equation}
where $x$ is the transverse spatial coordinate and $s$=1 ($s$=$-1$) 
corresponds to a focusing (defocusing) nonlinearity. 
The evolution coordinate $z$ can be time, as for Bose-Einstein 
Condensates (BEC's), or the spatial propagation coordinate, as for 
optical beams.
We consider only symmetric spatial response functions that are positive
definite and (without loss of generality) obey the normalization condition
\begin{equation}
  \int_{-\infty }^\infty R(x) dx = 1.
  \label{normalization}
\end{equation}
Thus we exclude asymmetric effects, such as those generated by asymmetric 
temporal response functions (with $x$ being time), as in the case of the 
Raman effect on optical pulses \cite{Wyller}.

In nonlinear optics Eq.~(\ref{nonlocal}) represents a general phenomenological 
model for media in which the nonlinear refractive index change (or 
polarization) induced by an optical beam is determined by some kind of a 
transport process.  
It may include, e.g., heat conduction in materials with a thermal 
nonlinearity \cite{thermal1,thermal2,thermal3} or diffusion of 
molecules or atoms accompanying nonlinear light propagation in atomic 
vapours \cite{suter}.  Nonlocality also accompanies the propagation of 
waves in plasma \cite{cusp,litvak,df,litvak75,juul}, and a nonlocal response 
in the form (\ref{nonlocal}) appears naturally as a result of many body 
interaction processes in the description of Bose-Einstein condensates 
\cite{bose}.  

\begin{figure}
  \epsfig{file=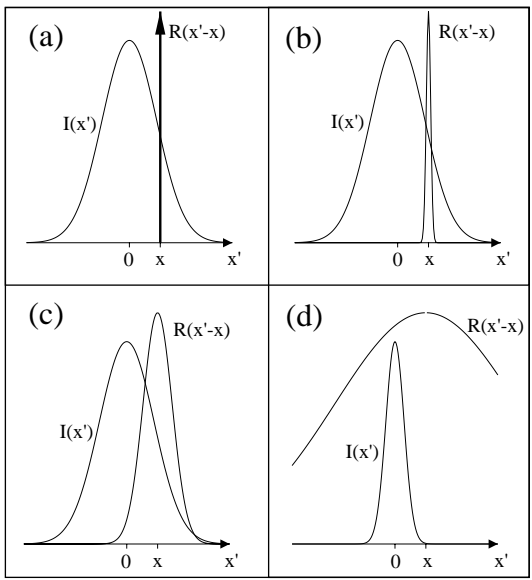,width=7cm, angle=0}
  \vspace{1ex}
  \caption{Different degrees of nonlocality, as given by the width of 
  the response function $R(x)$ and the intensity profile $I(x)$. 
  Shown is the local (a), the weakly nonlocal (b), the general (c),
  and the strongly nonlocal (d) response.} 
  \label{cases} 
\end{figure}

The width of the response function $R(x)$ relative to the width of the 
intensity profile $I(x,z)$ determines the degree of nonlocality, as
illustrated in Fig.~\ref{cases}. In the limit of a singular response, 
$R(x)$=$\delta(x)$ [see Fig.~\ref{cases}(a)], the nonlinearity is a 
local function of the intensity, 
\begin{equation}
  \Delta n = s I,
  \label{local}
\end{equation}
i.e.~the refractive index change at a given point is solely determined 
by the wave amplitude at that very point. For this well-known Kerr 
nonlinearity, which appears in all areas of physics, MI depends only 
on the sign $s$ \cite{mi-nlo1}.

With increasing width of $R(x)$ the intensity in the vicinity of the point 
$x$ also contributes to the index change at that point.
For a weak nonlocality, when the width of $R(x)$ is finite but still small 
compared to the width of $I(x,z)$ [see Fig.~\ref{cases}(b)], one can expand 
$I(x',z)$ around $x'$=$x$ in (\ref{nonlocal}) and obtain the simplified model
\begin{equation}
  \Delta n = s(I+\gamma\partial_x^2I), \quad
  \gamma   = \frac{1}{2}\int_{-\infty}^\infty x^2R(x)dx,
  \label{weaknonlocal}
\end{equation}
where $\gamma$ is a small positive parameter.
This diffusion type model of the nonlocal nonlinearity is a model in 
its own right in plasma physics, where $\gamma$ can take any sign 
\cite{cusp,litvak,df}. 
It was also applied to BEC's \cite{Parola98}, nonlinear optics 
\cite{wk-ob} and in the continuum limit of the theory of energy transfer in biomolecular systems \cite{molecular}.
In contrast to the local Kerr limit the MI now depends not only on the
sign $s$, but also on the intensity of the plane wave \cite{litvak}.

In the limit of a strongly nonlocal response, where $R(x)$ is much broader
than the intensity profile [see Fig.~\ref{cases}(d)], one can expand the 
response function around $x'$=$x$ in (\ref{nonlocal}) and obtain the simple 
{\em linear} model 
\begin{equation}
  \Delta n = s(c_0+c_1x+c_2x^2),
  \label{strongnonlocal}
\end{equation}
where $c_{0-2}$ are constants. Since this model is linear
all plane waves are stable, - there is no MI. 
The evolution of optical beams in such a medium was considered in the case 
when $c_1$=0 \cite{Snyder97}, which requires the center of mass of the beam 
to be always zero.

Modulational instability has thus been studied in the different limits.
However, in the general case (\ref{nonlocal}) the problem of MI has only
been studied for a few particular cases of the response function $R(x)$.
Here we present an analytical study of the general case with arbitrary 
profile $R(x)$ and confirm the results for three specific examples. 
We further show that the MI results of the weakly nonlocal model 
(\ref{weaknonlocal}) lead to erroneous conclusions when 
used to predict the behavior  of the full  model (\ref{nonlocal}).

\section{General Theory}

We will consider an optical beam propagating along the $z$-axis, with
the scalar amplitude of the electric field expressed as
\begin{equation}
  E(x,z)=\psi(x,z)\exp(iKz-i\Omega t) + c.c.,
  \label{field}
\end{equation}
where $K$ is the wavenumber, $\Omega$ is the optical frequency and
$\psi(x,z)$ is the slowly varying amplitude.
Substituting Eqs.~(\ref{nonlocal}) and (\ref{field}) into Maxwell's
equations we obtain the nonlocal nonlinear Schr\"{o}dinger (NLS) equation
\begin{equation}
  i\partial_z\psi + \frac{1}{2}\partial_x^2\psi +
  s\psi \int_{-\infty}^{\infty}R(x'-x)I(x',z) dx' = 0,
  \label{nonlocalNLS}
\end{equation}
where $I(x,z)$=$|\psi(x,z)|^2$ is the intensity of the beam. The model
(\ref{nonlocalNLS}) permits plane wave solutions of the form
\begin{equation}
  \psi(x,z) = \sqrt{\rho _0}\exp(ik_0x-i\omega_0z), \quad \rho_0>0,
  \label{planewave}
\end{equation}
where $\rho_0$, $k_0$, and $\omega_0$ are linked through the nonlinear
dispersion relation
\begin{equation}
  \omega_0 = \frac{1}{2}k_0^2 - s\rho_0,
  \label{nonlindisp}
\end{equation}
which is the same as for the standard local NLS equation
\begin{equation}
  i\partial_z\psi + \frac{1}{2}\partial_x^2\psi + s|\psi|^2\psi = 0.
  \label{NLS}
\end{equation}
Next, let us carry out a linear stability analysis of the plane wave
solutions (\ref{planewave}). Assume that
\begin{equation}
  \psi(x,z) = [\sqrt{\rho_0}+a_1(x,z)] \exp(ik_0x-i\omega_0z),
\end{equation}
where $a_1(x,z)$ is a small complex perturbation.
Inserting this expression into the nonlocal NLS equation (\ref{nonlocalNLS})
and linearizing around the solution (\ref{planewave}) yields the evolution
equation for the perturbation
\begin{eqnarray}
  & & i\partial_\tau a_1 + \frac{1}{2}\partial_\xi^2 a_1  \nonumber \\
  & & + 2s\rho_0\int_{-\infty }^{\infty}R(\xi'-\xi){\rm Re}
\{a_1(\xi',\tau)\} dx'=0.
\end{eqnarray}
In deriving this equation we have used the nonlinear dispersion relation
(\ref{nonlindisp}) and gone into a coordinate frame moving with the group
velocity $c_g=d\omega_0/dk_0=k_0$, i.e.
\begin{equation}
  \tau = z, \quad \xi = x-c_gz
\end{equation}
Decomposing the perturbation into real and imaginary parts, $a_1=u+iv$,
we obtain two coupled equations
\begin{equation}
  \begin{array}{c}
    \partial_\tau u + \frac{1}{2}\partial_\xi^2 v = 0, \\ \\
    \partial_\tau v - \frac{1}{2}\partial_\xi^2 u - 2s\rho_0
    \int_{-\infty}^{\infty} R(\xi'-\xi) u(\xi',\tau) d\xi' = 0.
    \end{array}
\end{equation}
By introducing the Fourier transforms
\begin{eqnarray}
  \widehat{u}(k,\tau) &=& \int_{-\infty }^{\infty} u(\xi,\tau)
                          \exp[ik\xi] d\xi, \nonumber\\
  \widehat{v}(k,\tau) &=& \int_{-\infty }^{\infty} v(\xi,\tau)
                          \exp[ik\xi] d\xi, \nonumber \\
  \widehat{R}(k)      &=& \int_{-\infty }^{\infty} R(\xi)
                          \exp[ik\xi] d\xi,
\end{eqnarray}
and exploiting the convolution theorem for Fourier transforms, the
linearized system is converted to a set of ordinary differential equations
in $k$-space
\begin{equation}
  \begin{array}{c}
    \partial_\tau \widehat{u} - \frac{1}{2}k^{2}\widehat{v} = 0, \\ \\
    \partial_\tau \widehat{v} + \frac{1}{2}k^{2}\widehat{u} -
    2s\rho_0\widehat{R}\,\widehat{u} = 0,
  \end{array}
\end{equation}
which can be written in the compact matrix form
\begin{equation}
  \partial_\tau \underline{X} = \underline{\underline{A}} \; \underline{X},
\end{equation}
where the vector $\underline{X}$ and matrix $\underline{\underline{A}}$ are
defined as
\begin{equation}
  \underline{X}=\left[
  \begin{array}{c}
    \widehat{u} \\
    \widehat{v}
  \end{array}
  \right] , \; \underline{\underline{A}}=\left[
  \begin{array}{cc}
    0 & \frac{1}{2}k^{2} \\
    -\frac{1}{2}k^2+2s\rho_0\widehat{R}(k) & 0
  \end{array}
  \right].
\end{equation}
The eigenvalues $\lambda $ of the matrix $\underline{\underline{A}}$ are
given by
\begin{equation}
  \lambda^2 = -k^2\rho_0\left[\alpha k^2-s\widehat{R}(k) \right],
  \label{generaleigenvalue}
\end{equation}
where we have defined the parameter $\alpha$ as
\begin{equation}
  \alpha = 1/(4\rho_0)
\end{equation}
Notice that if the response function $R(x)$ is real and symmetric then so
is the Fourier spectrum of $R(x)$, i.e. $\widehat{R}(k) = \widehat{R}^*(k) = 
\widehat{R}(-k)$.

The general eigenvalue equation (\ref{generaleigenvalue}) constitutes the
basis of our further study. First of all we notice that $\widehat{R}(0)$=1,
since the response function is normalized according to
Eq.~(\ref{normalization}).
For a focusing nonlinearity ($s$=$+1$) we therefore always have, by continuity,
that $\alpha k^2-s\widehat{R}(k)<0$, and thus $\lambda^2>0$, in a certain
wavenumber band symmetrically centered about the origin, where $k$ is
sufficiently small.
It follows that {\em we always will have (long wave) MI in the focusing case,
independently of the details in the behavior of the response function}.

In contrast, in the defocusing case ($s$=$-1$), the stability properties
depend in a sensitive way on the response function. Thus the nonlocality
could possibly lead to MI, which would never be present in the defocusing
local NLS equation (\ref{NLS}).

The well-known modulational instability (stability) result for the standard
local NLS equation (\ref{NLS}) is easily recovered from the general eigenvalue
equation (\ref{generaleigenvalue}) by setting $R(x)$=$\delta(x)$, where
$\delta(x)$ is the Dirac delta function: We get
\begin{equation}
  \lambda^2 = -k^2\rho_0(\alpha k^2-s)
\end{equation}
where $s=+1$ $(s=-1)$ yields instability (stability).

\section{Weakly nonlocal limit}

The MI properties in the local limit (\ref{local}), described by the NLS 
equation (\ref{NLS}), are well-known and the strongly nonlocal limit 
(\ref{strongnonlocal}) is linear, so obviously it does not display MI.
Here we briefly consider the interesting regime of a 
weak nonlocality, i.e.,~the situation when the typical width of the response 
kernel is small compared to the characteristic length of the wave modulation 
(see Fig.~\ref{cases}).  
In this case the nonlinearity has the form (\ref{weaknonlocal}) and 
the NLS equation (\ref{nonlocalNLS}) reads \cite{Parola98,wk-ob}
\begin{equation}
  i\partial_z\psi + \frac{1}{2}\partial_x^2\psi + 
  s(|\psi|^2 +\gamma\partial_x^2|\psi|^2)\psi = 0,
  \label{NLS-weak}
\end{equation}
where the nonlocality or diffusion parameter $\gamma$ is defined in 
Eq.~(\ref{weaknonlocal}).
Equation (\ref{NLS-weak}) has been discussed in the literature in the 
context of plasma physics \cite{df} and the continuum limit of discrete 
molecular structures \cite{molecular,nakamura} and has been shown to posses 
bright and dark soliton solutions. Their exact analytical soliton 
form was found recently \cite{wk-ob}.
Importantly, $\gamma$ is small and positive when Eq.~(\ref{NLS-weak})
represents the weakly nonlocal limit of the general nonlocal model
(\ref{nonlocalNLS}), whereas $\gamma$ can take any sign in plasma physics, 
where Eq.~(\ref{NLS-weak}) is a model in its own right.

The instability condition for the weak nonlocality is most easily obtained 
by Taylor expanding the spectrum of the response kernel about $k$=0 and only 
retain the two lowest order nontrivial terms in the expansion, i.e.
\begin{equation}
  \widehat{R}(k) \simeq 1-\gamma k^2,
  \label{R-weak}
\end{equation}
where obviously $\gamma k^2\ll1$ must be fulfilled. The eigenvalue equation 
then reads 
\begin{equation}
  \lambda^2 = -k^2\rho_0[(\alpha+s\gamma)k^2-s].
  \label{lamb-weak}
\end{equation}
This result does not depend on the detailed behavior of the response 
kernel, as should be expected. 

The eigenvalue Eq.~(\ref{lamb-weak}) clearly shows that MI is always present 
for a focusing nonlinearity ($s$=$+1$), since for sufficiently small 
wavenumbers $\lambda^2>0$ for all values of $\alpha$ and $\gamma$.
This is in correspondance with the results of the general nonlocal model 
(\ref{nonlocalNLS}).

More interesting is the defocusing case ($s$=$-1$), where there is
a critical value $\rho_0^{\rm cr}$ of the intensity,
\begin{equation}
  \rho_0^{\rm cr} = 1/(4\gamma),
  \label{rho-cr}
\end{equation}
which depends on the degree of nonlocality through the parameter $\gamma$,
and corresponds to $\alpha=\gamma$.
When $\gamma$ is positive (and $s=-1$) low intensity plane waves with 
$\rho_0\le \rho_0^{\rm cr}$ are modulationally stable, whereas MI appears 
at high intensities, $\rho_0>\rho_0^{\rm cr}$, despite the nonlinearity 
being defocusing. 
This MI is short-wave, appearing for pertubations with wavenumbers exceeding 
the threshold value
\begin{equation}
  k^{\rm cr} = 1/\sqrt{|\alpha-\gamma|}.
  \label{k-cr}
\end{equation}
The stability criterion $\rho_0<\rho_0^{\rm cr}$ corresponds exactly to the
criterion for existence of dark soliton solutions in this case \cite{wk-ob}.
The existence of MI in the weakly nonlocal model is in sharp contrast to the 
local defocusing Kerr nonlinearity, for which there is no MI.

In the case $\gamma<0$ in Eq.(\ref{NLS-weak}), relevant in plasma physics, nonlocality does not
affect the MI properties: For a focusing (defocusing) nonlinearity all 
plane waves are modulationally unstable (stable).
In Table.\,I we have summarized the MI results of the weak nonlocality model.

\begin{minipage}[t]{8cm}
\begin{table}
\begin{tabular}{|c|c|c|}
   \hspace{15mm} & \hspace{5mm} s=1 \hspace{5mm} & \hspace{5mm} s=-1 
                   \hspace{5mm}  \\ \tableline
   $\gamma<0$ & MI &  STABILITY  \\ \tableline
   $\gamma>0$ & MI &  {$ \begin{array}{ccc}\mbox{STABILITY} & \mbox{for} 
                   & \rho_0< \rho_0^{\rm cr}\\ 
   \mbox{MI} & \mbox{for} & \rho_0>\rho_0^{\rm cr} \end{array}$} 
\end{tabular} \vspace{1ex}
\caption{Modulational stability properties of the weakly nonlocal
model (\ref{NLS-weak}).}
\end{table}
\end{minipage}

\begin{figure}[H]
  \epsfig{file=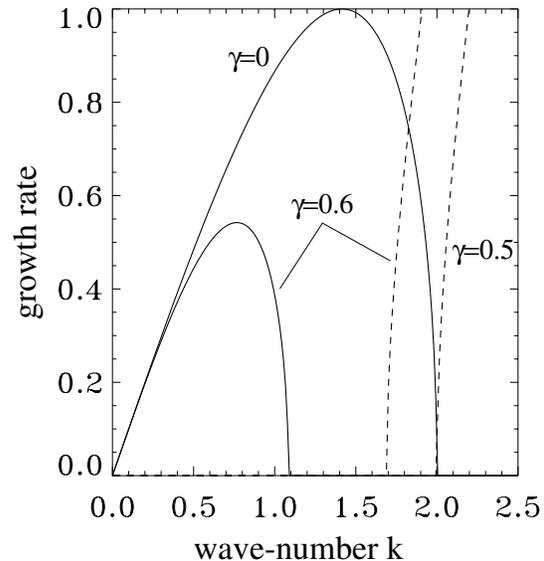,width=7cm,angle=0.0}
  \vspace{1ex}
  \caption{Growth rate vs.~wavenumber for plane waves with intensity 
  $\rho_0$=1 in media with a weakly nonlocal focusing (s=1, solid lines) 
  and defocusing (s=-1, dashed lines) nonlinearity.}
  \label{MIbands-weak-gampos}
\end{figure}

So only for defocusing media with $\gamma>0$ does a weak nonlocality
change the MI properties in terms of existence, as compared to the 
local NLS limit.
However, nonlocality does always affect the structure of the MI gain 
spectrum, as illustrated in Fig.~\ref{MIbands-weak-gampos} for $\gamma>0$.
In the focusing case Fig.~\ref{MIbands-weak-gampos} clearly shows how the 
nonlocality tends to suppress MI by decreasing both the maximum gain and
the gain bandwidth. 
In contrast, in the defocusing case, the {\em nonlocality promotes MI}, - the 
higher the degree of nonlocality $\gamma$ the lower the threshold value 
of the wavenumber, above which the instability develops. 

\begin{figure}
  \vspace{-0ex}
  \epsfig{file=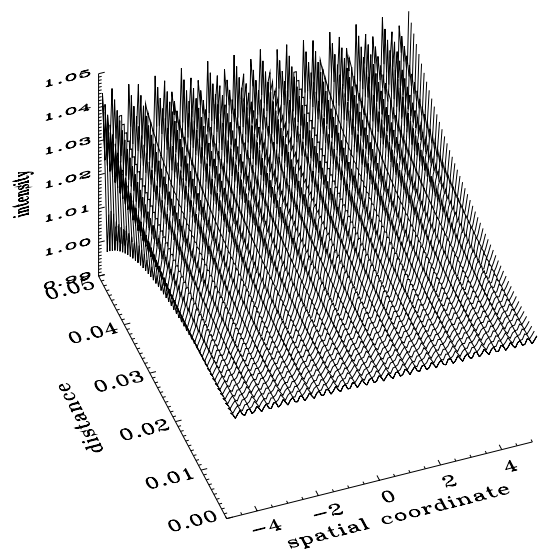,width=7cm,angle=0}
  \vspace{1ex}
  \caption{Development of MI of a plane wave with intensity $\rho_0$=1 
  in a weakly nonlocal defocusing (s=-1) medium. The initial amplitude is 
  modulated with the wave vector $k$=$7\pi$.}
  \label{5} 
\end{figure} 

In Fig.~3 we show the development of MI of a plane wave in a self-defocusing 
weakly nonlocal medium, as obtained by numerical integration of the NLS 
equation (\ref{NLS-weak}). As an initial condition we used a plane wave 
with periodically perturbed amplitude.
It is evident that this perturbation grows as the wave propagates.


Although Eqs.(\ref{rho-cr}) and (\ref{k-cr}) predict MI in the defocusing 
case it is important to stress that this result does not actually apply to 
the case of weak nonlocality.  
In deriving the model (\ref{NLS-weak}) and the expansion (\ref{R-weak}) 
it was assumed that
\begin{equation}
  \label{pert-req}
  \gamma k^2\ll 1,
\end{equation}
which means that the width of the response function must be small compared 
to the characteristic length-scale of the modulation.
However, from Eq.~(\ref{lamb-weak}) we get that MI occurs when
\begin{equation}
  (\alpha-\gamma)k^2 + 1 < 0,
\end{equation}
which can be rewritten as
\begin{equation}
  \gamma k^2 > 1+\alpha k^2.
\end{equation}
Since $\alpha=1/(4\rho_0)$ is positive this condition cannot be fulfilled 
without violating the perturbation requirement (\ref{pert-req}).
Hence, in the region of parameters where MI occurs (and dark solitons
do not exist), the weak nonlocality approximation breaks down. 
This is in accordance with the analysis of the full nonlocal model 
(\ref{nonlocalNLS}), which, as we show below, does not display MI for 
self-defocusing nonlinearities with a smooth nonlocal response function. 

It should be stressed however that the appearance of MI is a valid result in 
the situation when the weakly nonlocal model (\ref{nonlocalNLS}) is a model
in its own right, as for example for upper hybrid modes in plasma \cite{df}.

\section{Fully nonlocal response - examples}

We will now discuss the stability issue in the general nonlocal Kerr 
nonlinearity with an arbitrary degree of nonlocality by considering a 
few specific examples of the nonlocal response function.  

\subsection{Gaussian response function}

Let the response function of the nonlocal nonlinear medium have the following 
Gaussian form 
\begin{equation}
  R(x) = \frac{1}{\sigma\sqrt{\pi}}\exp\left[-\frac{x^{2}}{\sigma^2}\right].
  \label{gaussprofile}
\end{equation}
In this case $\gamma=\sigma^2/2$, the Fourier transform of the response 
function is given by
\begin{equation}
  \widehat{R}(k) =\exp\left[-\frac{1}{4}\sigma^2k^2\right],
\end{equation}
and the eigenvalue equation (\ref{generaleigenvalue}) reduces to
\begin{equation}
  \lambda^2 = -k^2\rho_0\left(\alpha k^2-s\exp \left[-\frac{1}{4}
  \sigma^2k^2\right]\right).
\end{equation}

For the defocusing case ($s=-1$) we find modulational stability for all 
values of the intensity $\rho_0$, in contrast to the behaviour predicted 
by the weakly nonlocal model (\ref{NLS-weak}), according to which plane 
waves become unstable if their intensity is higher than the critical 
intensity $\rho_0^{\rm cr}$. Thus, as expected, {\em this proves that the
weakly nonlocal model gives incorrect predictions of the MI properties
of the full general nonlocal model}.

\begin{figure}\vspace{-0ex}
  \epsfig{file=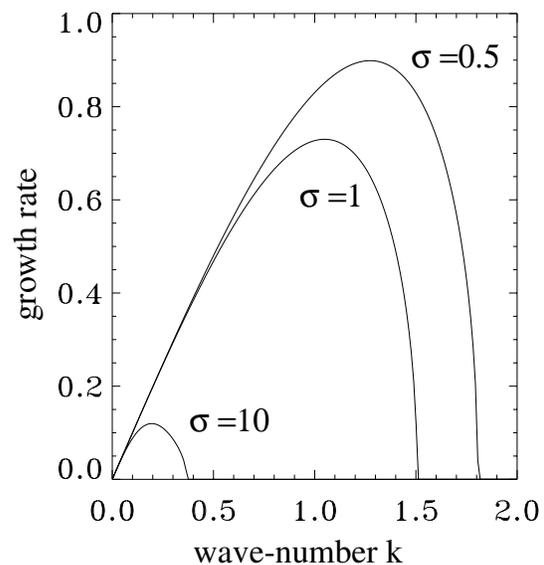,width=7cm,angle=0}
  \vspace{1ex}
  \caption{MI gain profiles for $\rho_0$=1 in self-focusing (s=1) nonlocal Kerr 
  media with a Gaussian response profile.} 
  \label{gain-gauss} 
\end{figure}

A simple analysis reveals finite band width MI in the focusing case 
($s=+1$) with a growth rate given by
\begin{equation}
  \left| \mbox{Re}\lambda \right| =\left| k\right| \sqrt{\rho _{0}}\sqrt{\exp 
  \left[ -\frac{1}{4}\sigma ^{2}k^{2}\right] -\alpha k^{2}}
\end{equation}
The dependence of the growth rate on the wave number and degree of 
nonlocality $\sigma$ is displayed in Fig.~\ref{gain-gauss}. 
Evidently, as the degree of the nonlocality increases the MI bandwidth 
shrinks and the maximum growth rate decreases. 
Thus nonlocality clearly tends to suppress MI in this case, as also 
predicted by the weakly nonlocal model.

\begin{figure}
  \vspace{-0ex}
  \epsfig{file=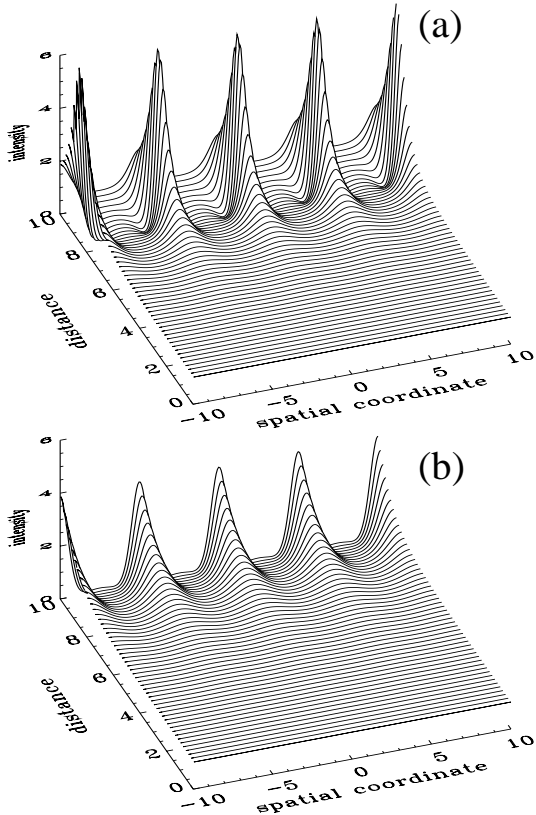,width=7cm,angle=0}
  \vspace{1ex}
  \caption{Development of MI of a unit amplitude plane wave in a nonlocal 
  self-focusing (s=1)medium with a Gaussian response function for (a) 
  $\sigma$=0.1 and (b) $\sigma$=1.0} 
  \label{dynam-gauss} 
\end{figure}

In Fig.~\ref{dynam-gauss} we demonstrate the influence of nonlocality on the
dynamic development of MI. 
To this end we integrated numerically the NLS equation (\ref{nonlocalNLS}) 
with the Gaussian response function (\ref{gaussprofile}). 
As an initial conditions we used a unit intensity plane wave with imposed 
weak periodic perturbation
\begin{equation}
  \psi(x,z=0) = 1+10^{-4}\cos(1.5x)
\end{equation}
It is evident that increasing the degree of nonlocality $\sigma$ 
suppresses the instability by lowering the gain. 

\subsection{Exponential response function}

Another interesting and important example involves the exponential response 
function 
\begin{equation}
  R(x) = \frac{1}{2\sigma}\exp\left[-\frac{|x|}{\sigma}\right],
  \label{diff-response}
\end{equation}
for which $\gamma=2\sigma^2$, with $\sigma$ again determining the degree
of nonlocality.
The importance of this particular response lies in the fact that it applies
exactly to media with a nonlinearity (i.e.~nonlinear refractive index 
$\Delta n(x)$) described by the diffusion equation
\begin{equation}
  -\sigma^2\frac{\partial^2\Delta n}{\partial x^2}+\Delta n = |\psi|^2.
  \label{diffusion}
\end{equation}
The equivalence between the nonlocal nonlinearity (\ref{nonlocal}) with 
the exponential response (\ref{diff-response}) and the diffusion model 
(\ref{diffusion}) is easily shown by Fourier transformation. 

\begin{figure}
  \epsfig{file=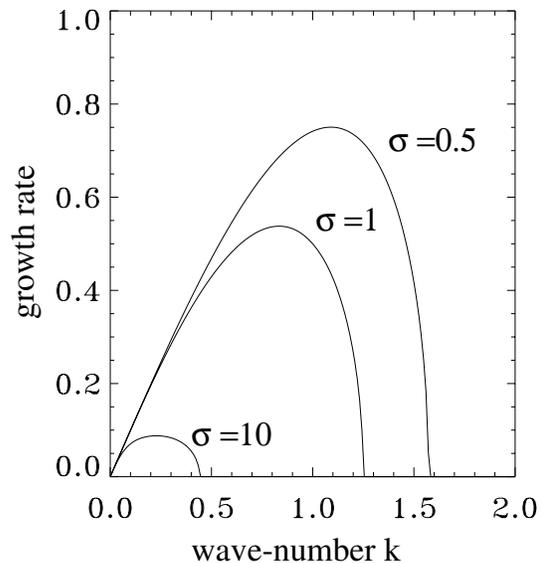,width=7cm,angle=0}
  \vspace{3ex}
  \caption{MI gain profiles for $\rho_0$=1 in nonlocal self-focusing 
  (s=1) media with an exponential response function.} 
  \label{gauss-fig}
\end{figure}

As discussed by Litvak et al.~\cite{litvak75} the model (\ref{diffusion}) 
describes the nonlinear thermal self-focusing of electromagnetic waves  
in a weakly ionized plasma. In this particular case $\Delta n$ corresponds 
to the change of the plasma temperature. From the Fourier transform
\begin{equation}
  \widehat{R}(k) = \frac{1}{1+\sigma^2k^2},
\end{equation} 
we obtain the eigenvalue equation
\begin{equation}
  \lambda^2 = -k^2\rho_0 \left[ \alpha k^2-s\frac{1}{1+(\sigma k)^2}\right],
  \label{lambda_diff}
\end{equation}  
which coincides with that derived in Ref.~\cite{litvak75} [Eq.~(4.1) in 
this reference]. The dispersion relation (\ref{lambda_diff}) imposes MI
properties similar to those found for the Gaussian response. 
In particular, it follows that in the self-defocusing regime ($s=-1$) the 
plane wave solution is {\em always} modulationally stable. On the other 
hand, in the self-focusing case plane wave develops transverse instability 
with the growth rate given by
\begin{equation}
  |\mbox{Re} \lambda| = |k|\sqrt{\rho_0}\sqrt{\frac{1}{1+(\sigma k)^2} -
  \alpha k^2}.
\end{equation}
In Fig.~(\ref{gauss-fig}) we show the MI growth rate versus the wave number 
for a few values of the nonlocality parameter $\sigma$. 
As for the Gaussian response the nonlocality tends to suppress MI by
narrowing the gain bandwidth and decreasing the maximum gain. 
However, only in the extreme limit of an infinitely nonlocal response 
($\sigma\rightarrow\infty$) does MI completely disappear, as the medium 
no longer exhibits any nonlinearity.

\subsection{Rectangular response function}

Finally we consider the rectangular response function 
\begin{equation}
  R(x) = \left\{ 
  \begin{array}{ccc}
    \frac{1}{2\sigma} & \mbox{for} & |x|\leq\sigma \\ \\ 
    0                 & \mbox{for} & |x|>\sigma
  \end{array}
  \right.
\end{equation}
for which $\gamma=\sigma^2/3$. From the Fourier transform 
\begin{equation}
  \widehat{R}(k) =\frac{\sin(k\sigma)}{k\sigma},
  \label{fourier-exp}
\end{equation}
we obtain the eigenvalue equation 
\begin{equation}
  \lambda^2 = -k^2\rho_0\left[ \alpha k^2-s\frac{\sin(k\sigma)}{k\sigma}
  \right].
\end{equation}
In the focusing case one again finds that there is always (for any value
of $\alpha $ and $\sigma$) a finite bandwidth MI gain band symmetrically 
located about the origin in $k$-space, i.e.~for $|k|\leq k_{\rm cr}$, where
\begin{equation}
 \alpha k_{\rm cr}^2 = \frac{\sin(k_{\rm cr}\sigma)}{k_{\rm cr}\sigma},
\end{equation}
This is illustrated in Fig.~7, where we show the gain profiles for several
degrees of nonlocality $\sigma$. 
Again MI is being suppressed by the nonlocality, but now additional MI 
bands appear for large values of $\sigma$ (see $\sigma$=20).

\begin{figure}
  \epsfig{file=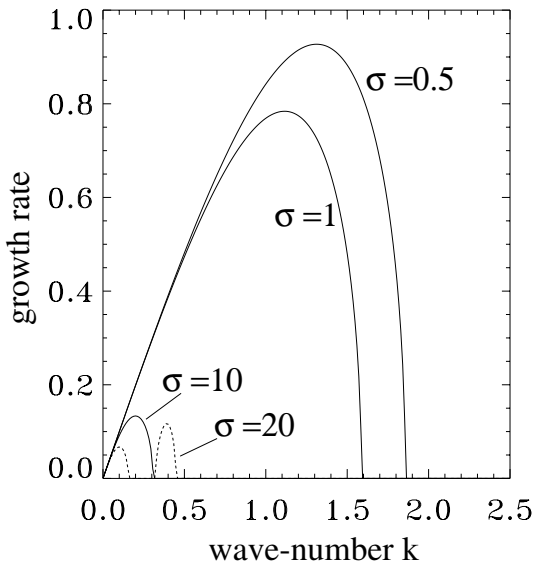,width=7cm,angle=0}
  \vspace{3ex}
  \caption{MI gain profiles for $\rho_0$=1 in self-focusing (s=1) nonlocal 
  media with a rectangular response function.} 
  \label{fig7} 
\end{figure} 

In the defocusing case a careful analysis reveals that for large and 
moderate values of the parameter $\alpha/\sigma^2$, i.e.~small and 
moderate values of $\rho_0\sigma^2$, plane waves are modulationally 
stable, while in the opposite regime one or more MI bands appear, as
shown in Fig.~8.
This instability is caused by the negative-sign bands of the 
sinc$(x)$-profile Fourier Transform (\ref{fourier-exp}), with the first 
instability band being in the region $\pi\le k\sigma\le 2\pi$.
More specifically the instability appears when the background intensity 
exceeds the threshold value 
\begin{equation}
  \rho_0^{\rm th} > 21.05/\sigma^2. 
\end{equation}

\begin{figure}
  \epsfig{file=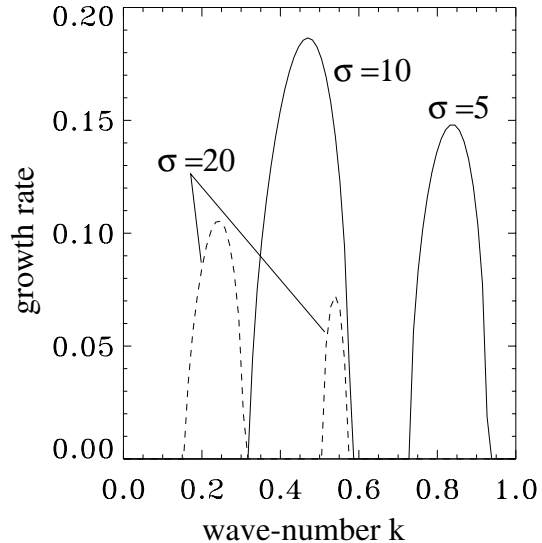,width=7cm,angle=0}
  \vspace{3ex}
  \caption{MI gain profiles for $\rho_0$=1 in self-defocusing (s=-1) nonlocal 
  media with a rectangular response function.} 
  \label{fig8} 
\end{figure} 

\section{Conclusions}

We studied the modulational stability properties of plane waves in 
nonlinear media with a general nonlocal Kerr nonlinearity, described 
by a nonlocal NLS equation. 
We derived an analytical expression for the growth rate of the instability 
for arbitrary response profiles, i.e.~for arbitrary degrees of nonlocality.   
We found that in the self-focusing case  the plane wave solution is always 
unstable, - the nonlocality tends to suppress the instability, independently 
of the form of the response function, but can never suppress is completely. 

In self-defocusing media the stability properties depend strongly on the
specific form of the nonlocal response. For any strictly positive 
continuous response profile all plane waves are stable for any degree of
nonlocality.However, exotic
profiles with jump-discontinuities like the rectangular response, may actually promote MI when the intensity is sufficiently high"

In the so-called weakly nonlocal limit (narrow response function compared
to the intensity profile) we find that plane waves become 
modulationally unstable if their intensity exceeds a certain critical value. 
However, we show that this surprising prediction of MI in the defocusing
case can never apply to the full nonlocal model with a smooth response 
profile. 
Thus one should be cautious with applying results of the simple weakly
nonlocal model to the real general case, in particular when studying
MI and dark solitons.

This work was supported by the Danish Technical Research Council
(STVF - Talent Grant 5600-00-0355), the Danish Natural Sciences
Foundation (SNF - grant 9903273), and the Graduate School in
Nonlinear Science (The Danish Research Academy).

\end{multicols}

\begin{thebibliography}{99}

\bibitem{mi-fluid}
  T.B. Benjamin and J.E. Feir, 
  J. Fluid. Mech. {\bf 27}, 417 (1967).

\bibitem{mi-plasma}
  A. Hasegawa, {\it Plasma Instabilities and Nonlinear Effects} 
  (Springer-Verlag, Heidelberg, 1975).

\bibitem{mi-nlo1}
  L.A. Ostrovskii, Sov. Phys. JETP {\bf 24}, 797 (1967).

\bibitem{mi-nlo2}
  V.I. Bespalov and V.I. Talanov, JETP Lett. {\bf 3}, 307 (1966);
  V.I. Karpman, JETP Lett. {\bf 6}, 277 (1967).

\bibitem{discrete-mi-molchain} 
  Yu.S. Kivshar and M. Peyrard,
  \pra {\bf 46}, 3198 (1992).

\bibitem{discrete-mi-array} 
  P.D. Miller and O. Bang,
  \pre {\bf 57}, 6038 (1998).

\bibitem{mi-dispersive}
  M.J. Potasek, 
  \ol {\bf 12}, 921 (1987)

\bibitem{mi-saturation}
  Yu.S. Kivshar, D. Anderson, and M. Lisak, 
  Phys. Scripta {\bf 48} 679 (1993).

\bibitem{mi-coherence}
  M. Soljacic, M. Segev, T. Coskun, D. Christodoulides, and A. Vishwanath, 
  \prl {\bf 84}, 467 (2000).

\bibitem{Wyller}
  J. Wyller, 
  "Nonlinear wavefields in optical fibres with finite time response 
  and amplification effects",
  Physica D (to appear).

\bibitem{thermal1}
  J.P. Gordon, R.C. Leite, R.S. Moore, S.P. Porto, and J.R. Whinnery, 
  J. Appl. Phys. {\bf 36}, 3 (1965).

\bibitem{thermal2}
  S. Akhmanov, D.P. Krindach, A.V. Migulin, A.P. Sukhorukov, 
  and R.V. Khokhlov,
  IEEE J. Quant. Electron. QE-{\bf 4}, 568 (1968).

\bibitem{thermal3}
  M. Horovitz, R. Daisy, O. Werner, and B. Fischer, 
  Opt. Lett. {\bf 17}, 475 (1992).

\bibitem{suter}
  D. Suter and T. Blasberg, 
  \pra {\bf 48}, 4583 (1993).

\bibitem{cusp}
  M.V. Porkolab and M.V. Goldman, 
  Phys. Fluids, {\bf 19}, 872 (1976).

\bibitem{litvak}
  A.G. Litvak and A.M. Sergeev, 
  JETP Lett. {\bf 27}, 517 (1978).

\bibitem{df}
  T.A. Davydova and A.I. Fishchuk,  
  Ukr. J. Phys. {\bf 40}, 487 (1995).

\bibitem{litvak75}
  A.G. Litvak, V.A. Mironov, G.M. Fraiman, and A.D. Yunakovskii,
  Sov. J. Plasma Phys., {\bf 1}, 31 (1975).

\bibitem{juul}
  H.L. Pecseli and J.J. Rasmussen, 
  Plasma Phys. {\bf 22}, 421 (1980).

\bibitem{bose}
  F. Dalfovo, S. Giorgini, L.P. Pitaevskii, and S.Stringari, 
  K.Goral, K.Rzazewski, 
  \pra {\bf 61} 051601R (2000); Rev. Mod. Phys. {\bf 71}, 463 (1999);
  V.M. Perez-Garcia, V.V. Konotop, and J.J. Garcia-Ripoll,
  \pre {\bf 62}, 4300 (2000).

\bibitem{Parola98}
  A. Parola, L. Salanich, and L. Reatto,
  \pra {\bf 57}, R3180 (1998).

\bibitem{molecular}
  X. Wang, D. W.Brown, K. Lindenberg, and B.J. West, 
  \pra {\bf 37}, 3557 (1988).

\bibitem{nakamura}
  A. Nakamura, 
  J. Phys. Soc. Japan {\bf 42}, 1824 (1977).

\bibitem{wk-ob}
  W. Krolikowski and O. Bang, 
  \pre {\bf 63}, 016610 (2001).

\bibitem{Snyder97} 
  A. Snyder and J. Mitchell, 
  Science {\bf 276}, 1538 (1997).










\end{thebibliography}
\end{document}